\documentclass[preprint,5p,times,twocolumn]{elsarticle}
\usepackage{epstopdf}
\usepackage{amsmath}
\usepackage{varioref}
\usepackage{xr-hyper}
\usepackage{hyperref}
\usepackage{multirow}
\usepackage{soul} 
\usepackage{color}
\usepackage{multirow}
\usepackage{hhline}
\usepackage{tabularx}
\usepackage[normalem]{ulem}

\hypersetup{colorlinks=true,citecolor=blue,linkcolor=blue,filecolor=blue,urlcolor=blue}

\def\be{\begin{equation}}
\def\ee{\end{equation}}
\def\ba{\begin{array}}
\def\ea{\end{array}}
\def\bea{\begin{eqnarray}}
\def\eea{\end{eqnarray}}

\usepackage{epsfig}

\usepackage{amssymb}

\journal{Physics Letters B}

\begin{document}

\begin{frontmatter}

\title{New parameterization of the effective field theory motivated
relativistic mean field model}

\author{Bharat Kumar$^{a,b}$}
\author{S. K. Singh$^{c}$}
\author{B. K. Agrawal$^{b,d}$}
\author{S. K. Patra$^{a,b}$\corref{SKP}}
\address{$^{a}$Institute of Physics, Sachivalaya Marg, Bhubaneswar - 751005, India.}
\address{$^{b}$Homi Bhabha National Institute, Anushakti Nagar, Mumbai - 400094, India.}
\address{$^{c}$Department of Physics, Indian Institute of Technology Roorkee, Roorkee 247667, India.}
\address{$^{d}$Saha Institute of Nuclear Physics, 1/AF,  Bidhannagar, Kolkata - 700064, India.}
\cortext[SKP]{Corresponding author}
\ead{patra@iopb.res.in}

\begin{abstract}
A new parameter set is generated for finite and infinite nuclear system within
the effective field theory motivated relativistic mean field (ERMF) formalism.
The isovector part of the ERMF model employed in the present  study
includes the coupling of nucleons to the $\delta$ and $\rho$ mesons
and the cross-coupling of $\rho$ mesons to the $\sigma$ and $\omega$
mesons. The results for the finite and infinite   nuclear systems
obtained using our parameter set are in harmony with the available
experimental data. 
We find the maximum mass of the neutron
star to be $ 2.03M_\odot$  and yet  a relatively  smaller radius
at the canonical mass, $ 12.69$ km,
as required by the available
data. 

\end{abstract}


\end{frontmatter}


\section{Introduction}
\label{sec1}

The nuclear physics inputs are essential in understanding the properties
of dense objects like neutron stars. The relativistic mean field models
based on the effective field theory (ERMF)  motivated Lagrangian density have
been instrumental in describing the neutron star properties, since,
the ERMF models enables one to readily include the contributions from various
degrees of freedoms such as hyperons, kaons and Bose condensates .
The model parameters are obtained by adjusting them  to
reproduce the experimental data on the bulk properties for a selected set
of finite nuclei. However, these parameterizations give remarkable results
for bulk properties such as binding energy, quadrupole moment, root
mean square radius not only for beta stable nuclei, but also for nuclei
away from the stability line \cite{gambhir90,estal01}.   However, the same 
model,  sometimes does not appropriately reproduce the behavior of
the symmetric nuclear matter and  pure neutron matter at supra-normal
densities as well as those for the pure neutron matter at the
sub-saturation densities.

The ERMF model usually includes the contributions from the self and
cross-couplings of isoscalar-scalar $\sigma$, isoscalar-vector $\omega$
and isovector-vector $\rho$ mesons.  The inclusion of various self and
cross-couplings makes the model flexible to accommodate various phenomena
associated with   the finite nuclei and neutron stars adequately without
compromising the quality of the fit to those data considered a
priory. 
For example, the self-coupling of $\sigma$ mesons remarkably reduces the 
nuclear matter incompressibility to the desired  values \cite{boguta77}.  
The cross-coupling of $\rho$ mesons with $\sigma$ or $\omega$
allows one to vary the neutron-skin thickness in a heavy nucleus like
$^{208}$Pb  over a wide range \cite{pika05,pika12}.  These cross-couplings 
are also essential to produce desired behavior for the equation of state 
of pure neutron matter.  Though, the effects are marginal, but, the quantitative
agreement with the available empirical informations call for them
\cite{pika12,Agrawal12}.

One may also consider the  contributions due to  the  couplings of
the meson field gradients  to the nucleons as well as the
tensor coupling of the mesons to the nucleons within the ERMF model
\cite{estal01}. These additional couplings are required from
the naturalness view point, but very often they are neglected. 
Only the parameterizations of the ERMF model in which the contributions from
gradient and tensor couplings of mesons to the nucleons considered
are the TM1$^*$, G1 and G2 \cite{estal01,furnstahl97}. However, these
parameterizations display some disconcerting features.   For instance,
the nuclear matter incompressibility and/or  the neutron-skin thickness
associated with the TM1$^*$, G1 and G2  parameter sets are  little too
large in view of their current estimates based on the measured values for
the isoscalar giant monopole and the isovector giant dipole resonances in
the $^{208}$Pb nucleus \cite{Garg, Rocamaza}.  The equation of state (EoS)
for the pure neutron matter at sub-saturation densities show noticeable
deviations with those calculated using realistic approaches.

In the present paper, our motivation is to construct a new parameter
set taking into account the multiple cross-couplings as well as the
addition of $\delta-$meson which are generally ignored. 
Our new parameterization is confronted with the EoS for the symmetric
and pure neutron matters available from diverse sources which indicate
that the proposed parameter set can be employed to model the finite nuclei
as well as the neutron stars.

The paper is organized as follows. Sec. \ref{sec2} is devoted to a brief
outline of the extended relativistic mean-field model. After getting our 
newly generated parameter set, we have calculated the bulk properties of finite
nuclei, nuclear matter and neutron star in Sec. \ref{sec3}. Finally,  the
concluding remarks are given in Sec. \ref{sec4}.

\section{The model}
\label{sec2}

Here, we start with the energy density functional for the ERMF model
which includes the contributions from $\delta-$meson to the
lowest order  and the  cross-coupling between $\omega$ and $\rho$ mesons
which were not considered earlier by TM1$^*$, G1 and G2 parameterizations. The energy
density functional can be written as 
\cite{estal01,furnstahl97,singh14},  

\begin{eqnarray}
{\cal E}({r})&=&  \sum_\alpha \varphi_\alpha^\dagger({r})
\Bigg\{ -i \mbox{\boldmath$\alpha$} \!\cdot\! \mbox{\boldmath$\nabla$}
+ \beta \left[M - \Phi (r) - \tau_3 D(r)\right]\nonumber \\
&&+ W({r})+ \frac{1}{2}\tau_3 R({r}) + \frac{1+\tau_3}{2} A ({r})\nonumber \\
& &
- \frac{i \beta\mbox{\boldmath$\alpha$}}{2M}\!\cdot\!
  \left (f_\omega \mbox{\boldmath$\nabla$} W({r})
  + \frac{1}{2}f_\rho\tau_3 \mbox{\boldmath$\nabla$} R({r})+\lambda  \mbox{\boldmath$\nabla$} A
 \right)\nonumber\\
&&
+\frac{1}{2 M^2} (\beta_\sigma+\beta_\omega \tau_3)\Delta A 
  \Bigg\} \varphi_\alpha (r)\nonumber\\
&& + \left ( \frac{1}{2}
  + \frac{\kappa_3}{3!}\frac{\Phi({r})}{M}
  + \frac{\kappa_4}{4!}\frac{\Phi^2({r})}{M^2}\right )
   \frac{m_s^2}{g_s^2} \Phi^2({r}) \nonumber \\
&& -  \frac{\zeta_0}{4!} \frac{1}{ g_\omega^2 } W^4 ({r})
+ \frac{1}{2g_s^2}\left( 1 +
\alpha_1\frac{\Phi({r})}{M}\right) \left(
\mbox{\boldmath $\nabla$}\Phi({r})\right)^2
 \nonumber \\
 &&- \frac{1}{2g_\omega^2}\left( 1 +\alpha_2\frac{\Phi({r})}{M}\right)
\left( \mbox{\boldmath $\nabla$} W({r})  \right)^2 \nonumber \\
&& - \frac{1}{2}\left(1 + \eta_1 \frac{\Phi({r})}{M} +
 \frac{\eta_2}{2} \frac{\Phi^2 ({r})}{M^2} \right)
  \frac{m_\omega^2}{g_\omega^2} W^2 ({r})\nonumber\\
&&- \frac{1}{2e^2} \left( \mbox{\boldmath $\nabla$} A({r})\right)^2
 - \frac{1}{2g_\rho^2} \left( \mbox{\boldmath $\nabla$} R({r})\right)^2
 \nonumber \\
&&- \frac{1}{2} \left( 1 + \eta_\rho \frac{\Phi({r})}{M} \right)
   \frac{m_\rho^2}{g_\rho^2} R^2({r})\nonumber\\
&&- \frac{\eta_{2\rho}}{4 M^2}\frac{{m_\rho}^2}{{g_\rho}^2}\left(R^{2}(r)\times W^{2}(r)\right) \nonumber \\
&&+\frac{1}{2 g_{\delta}^{2}}\left( \mbox{\boldmath $\nabla$} D({r})\right)^2
   +\frac{1}{2}\frac{ {  m_{\delta}}^2}{g_{\delta}^{2}}\left(D^{2}(r)\right)
\nonumber\\
&&
-\frac{1}{2 e^2}(\mbox{\boldmath $\nabla$}A)^2+\frac{1}{3 g_\gamma g_\omega}
A \Delta W +\frac{1}{g_\gamma g_\rho}A\Delta R.\;
\label{eq1}
\end{eqnarray}
The extended energy density functional with $\delta-$meson contains
the nucleons and other exchange mesons like $\sigma$, $\omega$ and 
$\rho-$meson and photon $A_{\mu}$. The effects of the $\delta-$meson to
the  bulk properties of finite nuclei  are nominal, but, the effects are
significant for the highly asymmetric dense nuclear matter. The $\delta-$
meson splits the effective masses of proton and neutron which influences
the production of $K^{+,-}$ and $\pi^{+}/\pi^{-}$ in the heavy ion
collision (HIC)~\cite{ferini05}.  Also, it increases the proton fraction
in $\beta-$stable matter and modifies the transport properties of neutron
star and heavy ion reaction \cite{chiu64,bahc65,lati91}.
Furthermore, in Eq.\ref{eq1}, the terms having $g_\gamma, \lambda,
  \beta_\sigma $ and $\beta_\omega$ are responsible for the effects related
with the electromagnetic structure of the pion and nucleon \cite{furnstahl97}.
 We need to get the constant $\lambda$ to reproduce the magnetic moments 
of the nuclei and is defined by

\begin{eqnarray}
\lambda=\frac{1}{2}\lambda_p (1+\tau_3)+\frac{1}{2}\lambda_n (1-\tau_3)
\end{eqnarray}
with $\lambda_p=1.793 $ and $\lambda_n=-1.913$ the anomalous magnetic moments
for the proton and neutron, respectively \cite{furnstahl97}.

Certainly, the pairing correlation plays an important role for open-shell
nuclei. The effect can not be ignored especially for heavy mass nuclei because 
the availability of quasi-particles states near the Fermi surface. 
The simple BCS approximation is an appropriate formalism for nuclei near the 
stability line. However, it breaks down for nuclei far away from it. 
The reason behind such anomaly is the number of protons/neutrons increases as it 
goes away from the stability valley. For such nuclei the Fermi level approaches 
zero and  the number of available levels above the Fermi surface decreases. 
In this situation, the particle-hole and pair excitations reach the continuum. 
To overcome this problem, the BCS formalism is modified 
\cite{meyer98,liotta20} in an approximate manner by including the quasibound 
states (i.e.,  states bound by their centrifugal-plus-Coulomb barrier). 
In this present calculations, we have used the quasibound-BCS approach as done
 in Ref. \cite{estal201} to take care of the pairing interaction.

\begin{table}[t]
\caption{The obtained new parameter set G3 along with NL3 \cite{lala97},
FSUGold2 \cite{fsu2}, FSUGarnet \cite{chen15} and G2 \cite{furnstahl97} sets 
are  listed. The
nucleon mass $M$ is 939.0 MeV.  All the coupling constants are
dimensionless, except $k_3$ which is in fm$^{-1}$.  The lower
portion of the table indicates the nuclear matter properties such as
binding energy per nucleon $\mathcal{E}_{0}$(MeV), saturation density
$\rho_{0}$(fm$^{-3}$), incompressibility coefficient for symmetric nuclear
matter $K_{\infty}$(MeV), effective mass ratio $M^*/M$, symmetry energy
$J$(MeV) and linear density dependence of the symmetry energy $L$(MeV). }
\scalebox{0.8}{
\begin{tabular}{cccccccccc}
\hline
\hline
\multicolumn{1}{c}{}
&\multicolumn{1}{c}{NL3}
&\multicolumn{1}{c}{FSUGold2}
&\multicolumn{1}{c}{FSUGarnet}
&\multicolumn{1}{c}{G2}
&\multicolumn{1}{c}{G3}\\
\hline
$m_{s}/M$  &  0.541  & 0.530 & 0.529&0.554  &    0.559  \\
$m_{\omega}/M$  &  0.833  & 0.833&0.833 & 0.832  &    0.832  \\
$m_{\rho}/M$  &  0.812 & 0.812 &0.812 & 0.820  &    0.820  \\
$m_{\delta}/M$   & 0.0  & 0.0 & 0.0& 0.0  &    1.043  \\
$g_{s}/4 \pi$  &  0.813  &  0.827&0.837 & 0.835  &    0.782  \\
$g_{\omega}/4 \pi$  &  1.024  &  1.079&1.091 & 1.016  &    0.923 \\
$g_{\rho}/4 \pi$  &  0.712  & 0.714 &1.121&  0.755  &   0.962   \\
$g_{\delta}/4 \pi$  &  0.0  & 0.0 & 0.0& 0.0  &    0.160  \\
$k_{3} $   &  1.465  & 1.231 &1.368&  3.247  &  2.606  \\
$k_{4}$  &  -5.688  & -0.205 & -1.397& 0.632  &  1.694   \\
$\zeta_{0}$  &  0.0  & 4.705 &4.410&  2.642  &  1.010    \\
$\eta_{1}$  &  0.0  & 0.0 &0.0&  0.650  &   0.424   \\
$\eta_{2}$  &  0.0  & 0.0 &0.0&  0.110  &   0.114   \\
$\eta_{\rho}$  &  0.0  & 0.0 &0.0&  0.390  &  0.645   \\
$\eta_{2\rho}$  &  0.0  & 0.401 &50.698 &  0.0  &   33.250   \\
$\alpha_{1}$  &  0.0  & 0.0 &0.0&  1.723  &   2.000  \\
$\alpha_{2}$  &  0.0  & 0.0 &0.0&  -1.580  &  -1.468  \\
$f_\omega/4$  &  0.0  &  0.0 &0.0& 0.173  &   0.220 \\
$f_\rho/4$  &  0.0  &  0.0 &0.0& 0.962  &  1.239 \\
$\beta_\sigma$  &  0.0  &  0.0 &0.0& -0.093  &-0.087   \\
$\beta_\omega$  &  0.0  &  0.0 &0.0& -0.460  &-0.484   \\
\hline
& &  &  &  &&   &&&\\
$\mathcal{E}_{0}$  &  -16.29  & -16.28&-16.23 &  -16.07  &    -16.02 \\
$\rho_{0}$  &  0.148  & 0.1505&0.1529 &  0.153  &   0.148   \\
$K_{\infty}$  &  271.5  & 238.0&229.5 &  215.0  &    243.9   \\
$M^*/M$  &  0.595  & 0.593 &0.578&  0.664  &   0.699   \\
$J$  &  37.40  & 37.62 &30.95&  36.4  &   31.8   \\
$L$  &  118.6  & 112.9 &51.04& 100.0  &  47.3   \\
\hline
\end{tabular}}
\label{table1}
\end{table}

\section{Results and Discussions}
\label{sec3}

We have calibarated the parameters of the energy density functional
as given by Eq.(\ref {eq1}).  The optimization of the energy
density functional is performed for a given set of fit data using
the simulated annealing method. This method allows one to search
for the best fit parameter in a given domain of the parameter
space.  The detailed procedure of the parameterization  is given in
Refs. \cite{Agrawal12,Agrawal05}.  We have fitted the parameters or the
coupling constants to the properties of few spherical nuclei together
with  some constraints on the properties of the nuclear matter at the
saturation density.  The experimental data for the binding energies
and the charge radii for $^{16}$O, $^{40}$Ca, $^{48}$Ca, $^{68}$Ni,
$^{90}$Zr, $^{100,132}$Sn and  $^{208}$Pb nuclei are used to fit
the model parameters. The  values of nuclear matter incompressibility
$K_{\infty}$ and symmetry energy coefficient $J$ are constrained within
210$-$245 MeV and 28$-$35 MeV respectively.  The parameter $\zeta_0$
corresponding to the self-coupling of $\omega$ mesons is allowed to
vary within $1.0- 1.5$ in order to ensure that the maximum neutron star
mass is $\sim 2M_\odot$. The obtained parameter set G3 along with other
successful parameterizations NL3 \cite{lala97}, FSUGold2 \cite{fsu2},
FSUGarnet \cite{chen15} and G2 \cite{furnstahl97} are compared
in Table \ref{table1}.  The NL3 is an old  parameter set which has been
popularly used.  It includes self-coupling terms only for  $\sigma$ mesons
and all the cross-coupling terms are ignored.  The FSUGold2 and FSUGarnet
on the other hand in addition includes cross-coupling between $\omega$ and
$\rho$ mesons as well as the self-coupling term for the $\omega$ meson.
The G2 parameter set includes all the terms present in Eq. (\ref{eq1})
except those corresponding to the $\delta-$meson and $\omega$-$\rho$
couplings.  A detailed account on the importance of various couplings
can be obtained in Refs. \cite{estal01,maza11,alam15,biswal15}.

The parameters, such as $\eta_{1}, \eta_{2}, \eta_{2\rho}, \alpha_{1},
\alpha_{2}, f_{\omega}$ have their own importance to explain various
properties of finite nuclei and nuclear matter. For instance, the surface
properties of finite nuclei is analyzed through non-linear interactions
of $\eta_{1}$ and $\eta_{2}$ as discussed in Ref. \cite{estal01}.  It is 
known that addition of the isovector $\delta-$meson softens the symmetry energy at
subsaturation densities and it stiffens the   EoS at high densities \cite{kubis97,singh14}.  The $\delta-$meson does not significantly modify
the properties of finite nuclei, but it affects the maximum mass of the
neutron-star and some other properties for highly asymmetric systems.
Though, relevance of most of these parameters has been pointed out
in Ref. \cite{estal01} but, a more quantitative version  along this direction,
such as the uncertainties on the parameters and the correlations among
the parameters, needs to be pursued within the covariance approach
\cite{fsu2,nik15}.  An appropriate   covariance analysis for the
model considered in the present work requires a set of fitting data
which includes large variety of nuclear and neutron star observables.
The parameter obtained in the present work will facilitate  such an
investigation.

\begin{table}[t]
\caption{The binding energy per nucleon B/A(MeV) , charge radius $R_c$
(fm) and neutron skin thickness R$_{n}$-R$_{p}$ (fm) for some close
shell nuclei compared with the NL3, FSUGold2, FSUGarnet, G2 and G3 with 
experimental data \cite{audi12,angeli13}.}
\scalebox{0.7}{
\begin{tabular}{cccccccccc}
\hline
\hline
\multicolumn{1}{c}{Nucleus}&
\multicolumn{1}{c}{Obs.}&
\multicolumn{1}{c}{Expt.}&
\multicolumn{1}{c}{NL3}&
\multicolumn{1}{c}{FSUGold2}&
\multicolumn{1}{c}{FSUGarnet}&
\multicolumn{1}{c}{G2}&
\multicolumn{1}{c}{G3}\\
\hline
$^{16}$O&B/A & 7.976 &7.917&7.862&7.876 & 7.952&8.037  \\
         & R$_{c}$& 2.699 & 2.714&2.694&2.690   & 2.718&2.707  \\
         & R$_{n}$-R$_{p}$ & - & -0.026&-0.026&-0.028  & -0.028&-0.028  \\
\\
$^{40}$Ca & B/A  & 8.551 & 8.540&8.527&8.528  &8.529 &8.561  \\
         & R$_{c}$  & 3.478 &  3.466&3.444&3.438  & 3.453&3.459  \\
         & R$_{n}$-R$_{p}$  & - & -0.046&-0.047&-0.051  & -0.049&-0.049  \\
\\
$^{48}$Ca & B/A  & 8.666 & 8.636&8.616&8.609  & 8.668&8.671  \\
         & R$_{c}$  & 3.477 & 3.443&3.420&3.426   & 3.439&3.466  \\
         & R$_{n}$-R$_{p}$  & - &  0.229&0.235&0.169 & 0.213&0.174  \\
\\
$^{68}$Ni & B/A  & 8.682 & 8.698&8.690&8.692  & 8.682&8.690  \\
         & R$_{c}$  & - & 3.870 & 3.846&3.861  &3.861&3.892  \\
         & R$_{n}$-R$_{p}$  & - &0.262 &0.268&0.184 &0.240&0.190  \\
\\ 
$^{90}$Zr & B/A  & 8.709 & 8.695&8.685&8.693  & 8.684&8.699  \\
         & R$_{c}$  & 4.269 & 4.253& 4.230&4.231  & 4.240&4.276  \\
         & R$_{n}$-R$_{p}$  & - & 0.115&0.118&0.065   & 0.102&0.068 \\
\\
$^{100}$Sn & B/A  & 8.258 & 8.301& 8.282&8.298   & 8.248&8.266 \\
         & R$_{c}$  & - & 4.469&4.453&4.426   &4.470&4.497  \\
         & R$_{n}$-R$_{p}$  & - & -0.073&-0.075&-0.078  & -0.079& -0.079 \\
\\

$^{132}$Sn & B/A  & 8.355 & 8.371&8.361&8.372  & 8.366&8.359 \\
         & R$_{c}$  & 4.709 & 4.697&4.679&4.687   & 4.690&4.732  \\
         & R$_{n}$-R$_{p}$  & - & 0.349&0.356&0.224   & 0.322&0.243 \\
\\ 
$^{208}$Pb & B/A  & 7.867 &7.885& 7.881 &7.902 & 7.853&7.863 \\
         & R$_{c}$  &5.501  & 5.509&5.491&5.496 & 5.498&5.541  \\
         & R$_{n}$-R$_{p}$  & - &   0.283& 0.288&0.162 & 0.256& 0.180 \\
\hline

\hline

\hline
\end{tabular}
\label{table2}}
\end{table}

\begin{figure}[!b]
        \includegraphics[width=1.0\columnwidth,height=7cm]{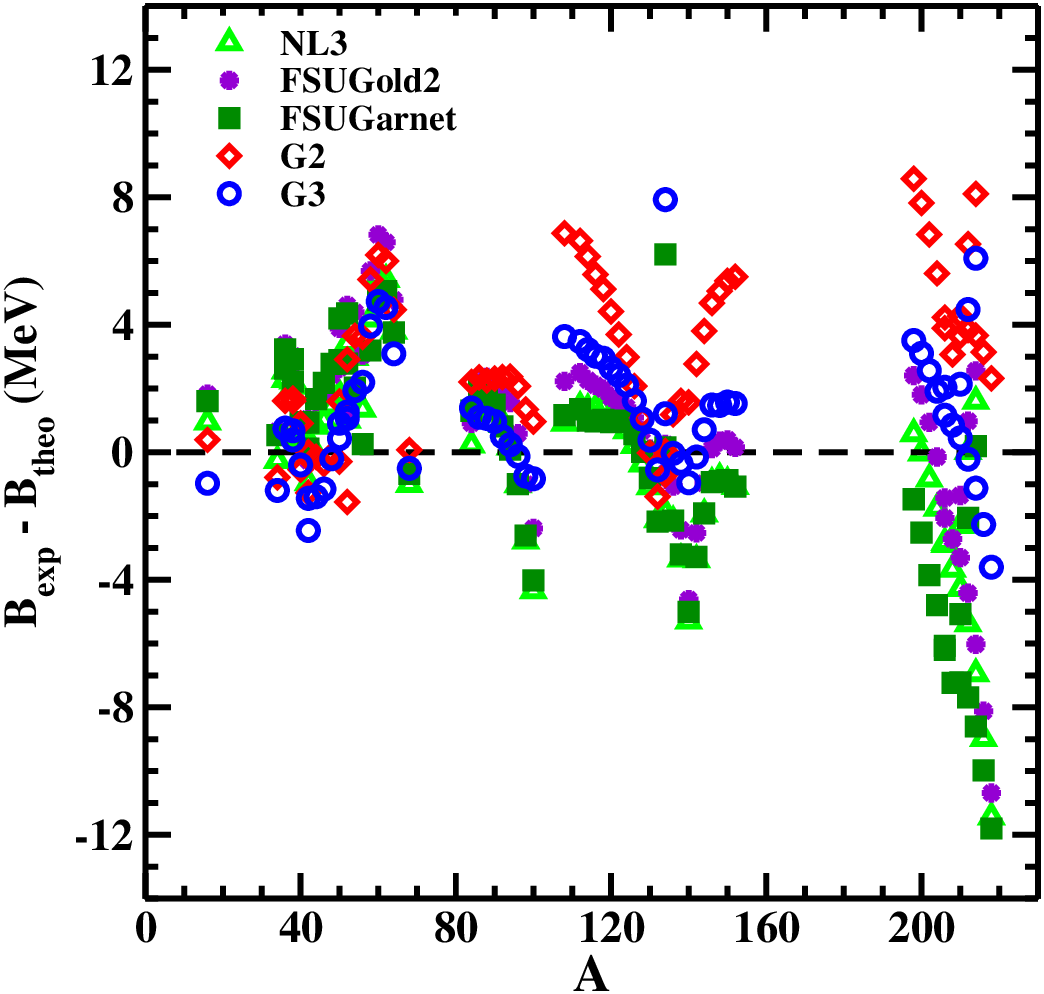}
\caption{(color online) Difference between experimental and 
theoretical binding energies as a function of mass numbers for
NL3 \cite{lala97}, FSUGold2 \cite{fsu2}, FSUGarnet \cite{chen15}, G2 \cite{furnstahl97} 
and G3 parameter sets. }
\label{be}
\end{figure}

The computed results  for NL3, FSUGold2, FSUGarnet, G2 and G3 are
listed  in Table \ref{table2}.  The binding energy per nucleon (B/A), root
mean square charge radius $R_{c}$ and neutron-skin thickness $R_{n}-R_{p}$
for some selected nuclei are compared with experimental data, wherever
available.  From the table, it seems that the predictive power of the new
set G3 for the nuclei considered in the fitting procedure  is as good as
for the  NL3, FSUGold2, FSUGarnet and G2 sets.  In Fig. \ref{be}
we plot the differences between the calculated and experimental binding
energies for 70 spherical nuclei \cite{klup09} obtained using different
parameter sets.  The triangles, stars, squares, diamonds and circles
are the results for the NL3, FSUGold2 , FSUGarnet, G2 and G3
parameterizations, respectively.  The above results affirm that G3 set
reproduces the experimental data better. The rms deviations  for the
binding energy as displayed in Fig. \ref{be} are  2.977, 3.062,
3.696, 3.827 and 2.308 MeV for NL3, FSUGold2, FSUGarnet, G2 and
G3 respectively. The rms error on the binding energy for G3 parameter
set is smaller in comparison to other parameter sets.

\begin{figure}[!b]
\includegraphics[width=1.0\columnwidth,height=7cm]{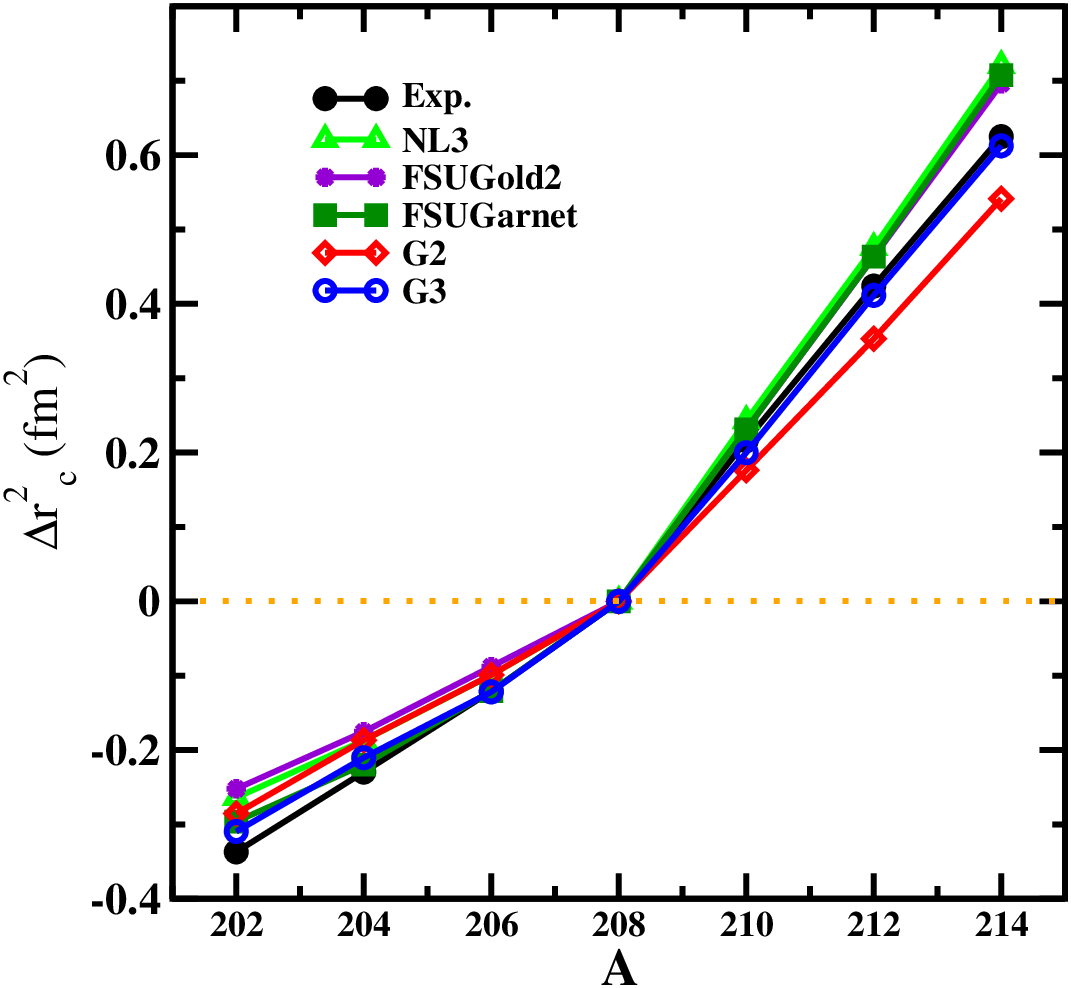}
\caption{(color online) The isotopic shift $\Delta{r_c}^2=
R_{c}^2(208)-R_{c}^2(A)$ (fm$^2$) of Pb isotopes taking $R_{c}$ of $^{208}$Pb
as the standard value.  Calculations with the NL3 \cite{lala97}, FSUGold2
\cite{fsu2}, FSUGarnet \cite{chen15}, G2 \cite{furnstahl97} and G3 parameter sets are compared.}
\label{shift}
\end{figure}

In Fig. \ref{shift}, the isotopic shift $\Delta{r}_c^2$ for Pb nucleus
is shown. The isotopic shift is defined as $\Delta{r}_c^2=R_{c}^2(A)-R_{c}^2(208)$
(fm$^2$), where $R_{c}^2(208)$ and $R_{c}^2(A)$ are the mean square radius
of $^{208}$Pb and  Pb isotopes having mass number A. From the figure,
one can see that $\Delta{r}_c^2$ increases with mass number monotonously
till A=208 ($\Delta{r}_c^2=0$ for $^{208}$Pb) and then gives a sudden
kink. It was first pointed by Sharma et al \cite{sharma93}, that the
non-relativistic parameterization fails to show this effect.  However,
this effect is well explained when a relativistic set like NL-SH
\cite{sharma93} is used. The NL3, FSUGold2 , FSUGarnet, G2 and G3
sets  also  appropriately predict this shift in Pb isotopes, but the
agreement with experimental data of the present parameter set G3  is
marginally better.

\begin{figure}[!b]
\includegraphics[width=1.0\columnwidth,height=7cm]{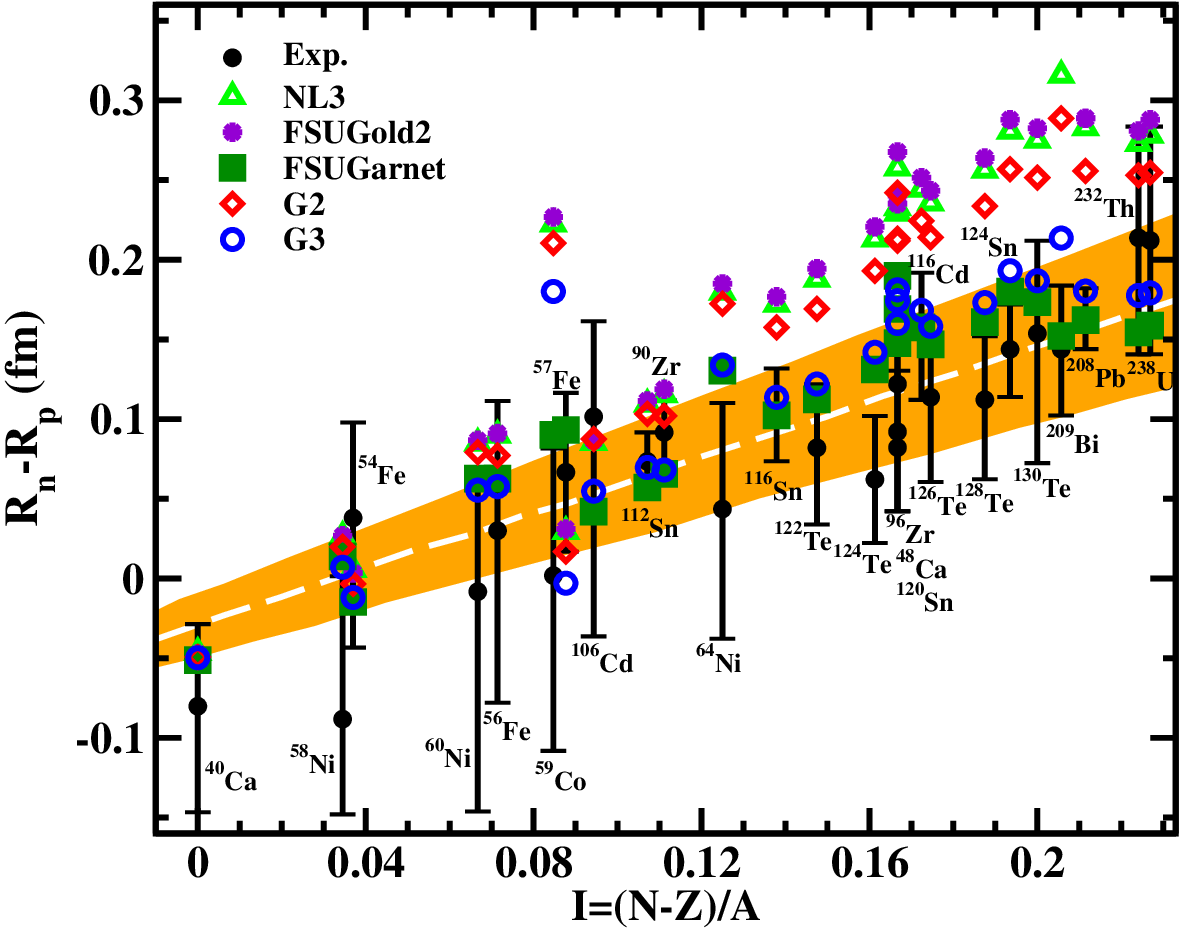}
\caption{(color online) The  difference in neutron and proton rms radii
$\Delta r_{np}$  obtained for NL3, FSUGold2, FSUGarnet, G2  and G3 are
plotted as a function of isospin asymmetric $I=(N-Z)/A$.  The experimental
data displayed are taken from \cite{thick,jast04}. The orange shaded
region represents Eq.  (\ref{eq:rnp}).
}
\label{skin}
\end{figure}

\begin{figure}[!b]
       \includegraphics[width=1.0\columnwidth,height=7cm]{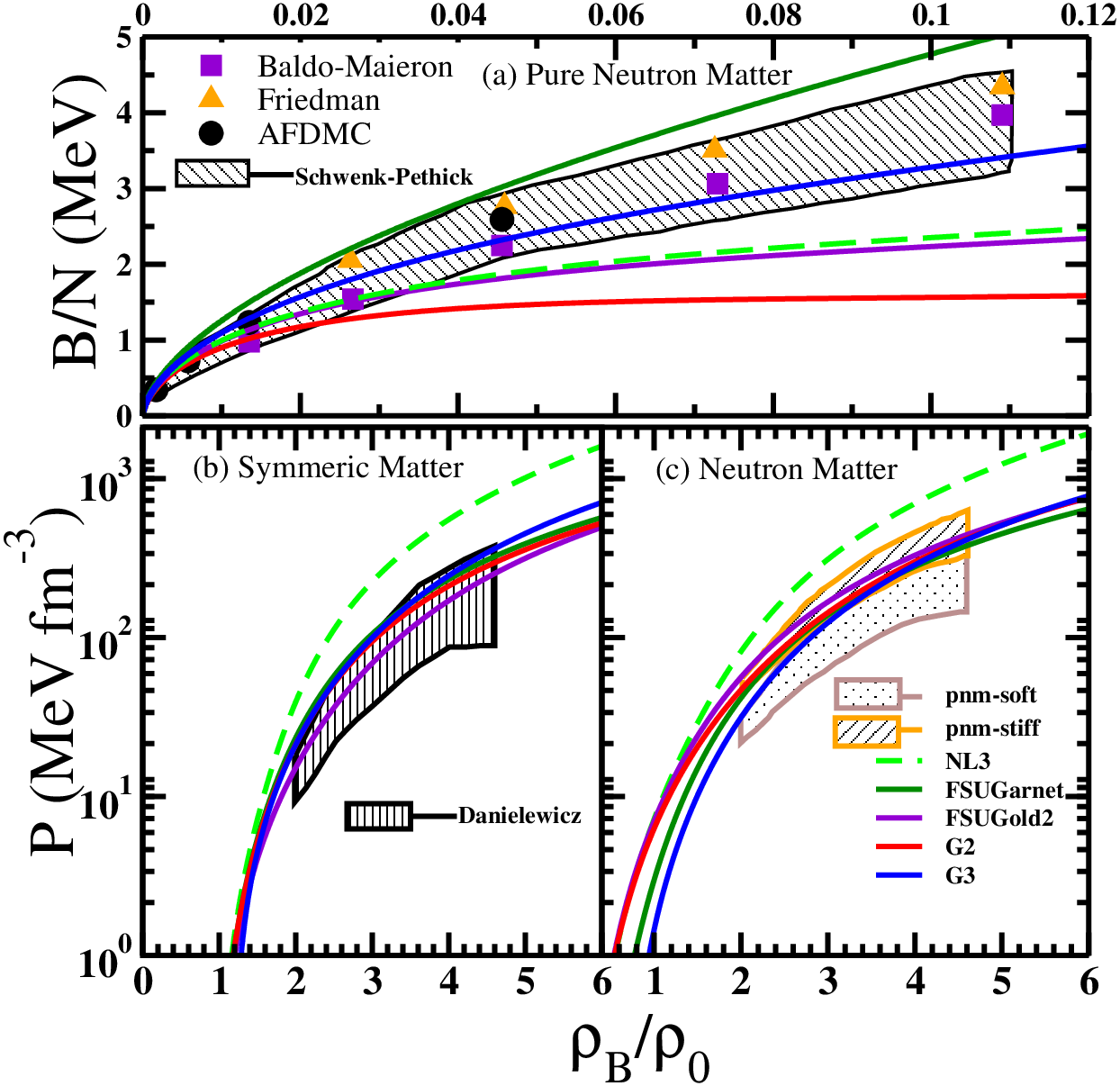} 
\caption{(color online) (a) The binding energy per neutron as a function
of neutron density for  G3 force is compared with other theoretical
calculations along with experimental data \cite{alex10,dutra12} for the
region of sub-saturation density. (b) and (c) are the pressure versus
baryon density for symmetric nuclear matter and pure neutron matter at
high densities, respectively.  The experimental data for higher density
region are  taken from \cite{daniel02}.  } \label{infinite} \end{figure}

The differences in the rms radii of  neutron and
proton distribution, $\Delta r_{np}=R_n-R_p$, the so-called neutron-skin
thickness  are plotted in Fig. \ref{skin}  for $^{40}$Ca to $^{238}$U for
NL3, FSUGold2, FSUGarnet, G2 and G3 parameter sets as a function of proton-neutron
asymmetry $I=(N-Z)/A$. The experimental data are also shown in the
figure. Trzci\'nska et al. extracted the neutron-skin thickness of 26
stable nuclei ranging from $^{40}$Ca to $^{238}$U from experiments done
with antiprotons at CERN \cite{thick,jast04}. Keen observation on the
data reveals more or less a linear dependence of neutron-skin thickness
on the relative neutron excess $I$ of nucleus. This can be fitted
by \cite{thick,jast04,xavier14}

\begin{eqnarray}
\Delta r_{np} = (0.90\pm0.15)I+(-0.03\pm0.02){\text fm }
\label{eq:rnp}
\end{eqnarray}

Eq.\ref{eq:rnp} is graphically represented in Fig. \ref{skin}
by the orange shaded region. 
Most of the $\Delta r_{np}$ obtained with NL3, FSUGold2,  and G2
overestimate the data and deviate from the 
shaded region. On the other hand, the $\Delta
r_{np}$ calculated using  G3 and FSUGarnet  lie in side  the shaded
region.  Interestingly, larger the asymmetry, more is the overestimation
of $\Delta r_{np}$ by NL3, FSUGold2, and G2 parameter sets. The $\Delta
r_{np}$,  calculated using G3 and FSUGarnet parameter sets are in harmony
with the experimental data.  The  overestimation of  $\Delta r_{np}$
for NL3, FSUGold2, and G2 parameter sets is due to the absence (or
negligible strength) of $\omega-\rho$ cross-coupling \cite{pika05}. This
term plays a crucial role in the determination of neutron distribution
radius $R_n$ without affecting much other properties of finite nuclei.
It is shown in Ref. \cite{brown00} that the derivative of neutron
matter EoS at a sub-saturation density is strongly correlated with the
$\Delta r_{np}$.  Further, one can readily verify that the behavior
of the neutron matter EoS should also depend on the incompressibility
coefficient $K_\infty$, since, the energy per nucleon for an asymmetric
matter can be decomposed into that for the symmetric nuclear matter and
the density dependent symmetry energy within a quadratic approximation.
Earlier parameterizations like TM1$^*$, G1 and G2 corresponding to
the Lagrangian density similar to the one used in the present work
yield higher values of $K_\infty$ and/or $J$.  In the present work,
we have attempted to improve this short coming and constructed the
force parameter G3 comprising $J=31.8$ MeV and $K_{\infty}=243.9$ MeV
(see Table \ref{table2}).

Our results for infinite symmetric nuclear and pure neutron matters are
shown in Fig \ref{infinite}.  The experimental data and predictions
of other theoretical approaches are also plotted for  comparison.
Fig. \ref{infinite}(a), displays the energy per neutron in pure neutron
matter at sub-saturation densities, which are encountered in finite
nuclei and in clusterization of nucleons. The  results for parameter sets
NL3 , FSUGold2, FSUGarnet and G2 deviate significantly from the
shaded region.  The non-relativistic forces labelled as Baldo-Maieron,
Friedman, AFDMC are designed for sub-saturated matter density , however,
they are not tested for the various  mass regions of finite nuclei.
The trend for the energy per neutron in pure neutron matter at low
densities obtained by our parameter set G3 passes well through the
shaded region.  The EoS for symmetric matter and pure neutron matter
are shown in Fig. \ref{infinite}(b) and Fig. \ref{infinite}(c),
respectively for various parameter sets. Except  the NL3, all other
EOSs for the SNM and PNM obtained using FSUGold2, FSUGarnet, G2 and
G3 are passing through the shaded region. 
Such a study by Arumugam
et al \cite{arumugam04} reported that EoS at high density
overestimates the experimental data in the absence of $\omega-$meson
self coupling and some cross-couplings.


Finally, we use our parameter
set to estimate the mass and radius of the static neutron star composed
of neutrons, protons, electrons and muons. The matter is assumed to be
in $\beta$-equilibrium and is charge neutral.  The contributions of the
crust EoS to the mass and the radius of the neutron star  for a given
central density are estimated using Ref.  \cite{Haensal17}.  It is shown
in Ref. \cite{Haensal17} that the mass and radius of the core for a given
central density together with the chemical potential at the core-crust
transition density  are enough to estimate reasonably well the thickness and
the mass of the crust.  We have used the core-crust transition density
to be 0.074 fm$^{-3}$ and the chemical potential at the transition point
to be 951.72 MeV, which altogether results in the total maximum 
mass,  $M_{\rm max}= 2.03 M_\odot$, and the corresponding radius 
$R_{\rm max}= 11.03$ km. The radius for the neutron star at the canonical 
mass is $R_{1.4} = 12.69$ km. The contribution due to crust to the total 
mass is $\sim 0.015M_\odot$ and those for the crust thickness at the 
maximum and the canonical masses are 0.39 and 1.06 km, respectively. These
values of the crust thicknesses are in harmony with the ones obtained in
Ref. \cite{Grill14} using appropriate EoSs for the inner and outer crusts.
Most of the relativistic mean-field models, in the absence of $\delta$-
mesons, which satisfy the observational constraint of $ 2M_\odot$ yield
$R_{1.4} > 13$ km \cite{Alam17}. The model DDH$\delta$ \cite{ddhd} which
includes the $\delta$-meson contributions yield $R_{1.4}$ similar to the
ones as presently obtained. Our value of $M_{\rm max}$ is consistent with
maximum mass so far observed for  neutron stars like PSR J1614-2230 has
$M=1.97\pm 0.04M_\odot$ \cite{demo10} and PSR J0348+0432 has $M=2.01\pm
0.04 M_\odot$ \cite{antoni13}.  The value of  $R_{1.4}= 12.69$  km is
also in good agreement with the empirical value $R_{1.4} = 10.7 - 13.1$
km, which is  consistent with the observational analysis and the host
of experimental data for finite nuclei \cite{Lattimer13}.

\section{Conclusions }
\label{sec4}

In conclusion, we  improve the existing parameterizations of the
ERMF model which includes couplings of the meson field gradients to
the nucleons and the tensor couplings of the mesons to the nucleons
in addition to the several self and cross-coupling terms. The nuclear
matter incompressibility coefficient  and/or symmetry energy coefficient
associated with  earlier parameterizations of such ERMF model were little
too large which has been taken care in our new parameter set G3. The
rms error on the total binding energy calculated for our parameter
set is noticeably smaller than the commonly used parameter sets NL3,
FSUGold2, FSUGarnet, and G2.  The neutron-skin thicknesses for our
parameterization calculated for nuclei  over a wide range of masses are
in harmony with the available experimental data. The neutron matter EoS at
sub-saturation densities for G3 parameter set show reasonable improvement over
other parameter sets considered. Our value for the maximum mass for the
neutron star is compatible with the measurements and the radius of the
neutron star with the canonical mass agree quite well with the empirical
values.  The smallness of $R_{1.4}$ for G3  parameter set in comparison to
those for  the earlier parametrization of the relativistic mean-field
models, which are compatible with the  observational constraint of
$2M_\odot$, is a desirable feature.

In the upcoming, we will perform a detailed covariance analysis for the
model used in the present work and  asses the uncertainties associated
with various parameters. An appropriate   covariance analysis of our model
requires a set of fitting data which includes large variety of nuclear
and neutron star observables.  The G3 parameter obtained in the present
work will facilitate  such an investigation. \\ 
\\
\\

{\bf { Acknowledgement:}}\\
S. K. Singh is supported by the Council of
Scientific and Industrial Research, Government of India, via Project
No. 03(1338)/15/EMR-II. \\
\\
\\



\begin{thebibliography}{00}


\bibitem{gambhir90}Y. K. Gambhir, P. Ring, and A. Thimet,
        Ann. Phys. (N.Y.) {\bf 198} (1990) 132.
\bibitem{estal01} M. Del Estal, M. Centelles, X. Vi\~nas and S. K. Patra,
        Phys. Rev. C {\bf 63} (2001) 024314.
\bibitem{boguta77}J. Boguta and A. R. Bodmer, Nucl. Phys. A {\bf 292} (1977)
413.

\bibitem{pika05} B. G. Todd-Rutel and J. Piekarewicz,Phys. Rev. Lett.
{\bf 95} (2005) 122501;
                 C. J. Horowitz and J. Piekarewicz, Phys. Rev. Lett.
{\bf 86} (2001) 5647;
                 Phys. Rev. C {\bf 64} (2001) 062802 (R).
\bibitem{pika12} J. Piekarewicz, B. K. Agrawal, G. Col\'o, W. Nazarewicz,
N. Paar, P. -G. Reinhard, X. Roca-Maza, and D. Vretenar,
Phys. Rev. C {\bf 85} (2012) 041302(R).

\bibitem{Agrawal12} B. K. Agrawal, A. Sulaksono,  P. -G. Reinhard,
Nucl. Phys. A {\bf 882} (2012) 1.

\bibitem{furnstahl97} R. J. Furnstahl, B. D. Serot and H. B. Tang,
        Nucl. Phys. A {\bf 598} (1996) 539;
        R. J. Furnstahl, B. D. Serot and H. B. Tang,
        Nucl. Phys. A {\bf 615} (1997) 441.
\bibitem{Garg} D. Patel, U. Garg, M. Fujiwara, H. Akimune, G. P. A. Berg,  
M. N. Harakeh, M. Itoh, T. Kawabata, K. Kawase, B. K. Nayak, T. Ohta, 
H. Ouchi, J. Piekarewicz, M. Uchida,  H.  P. Yoshida,  M. Yosoi, 
Phys. Lett. B {\bf 718} (2012) 447.
\bibitem{Rocamaza} X. Roca-Maza, X. Vi\~nas, M. Centelles, B. K. Agrawal, 
G. Col\'o, N. Paar, J. Piekarewicz, and D. Vretenar,  
Phys. Rev. C {\bf 92} (2015) 064304.
\bibitem{singh14}S. K. Singh, S. K. Biswal, M. Bhuyan, S. K. Patra,
        Phys. Rev. C {\bf 89} (2014) 044001.
\bibitem{ferini05} G. Ferini, M. Colonna, T. Gaitanos and M. Di Toro,
        Nucl. Phys. A {\bf 762} (2005) 147.
\bibitem{chiu64} H. Chiu and E. E. Salpeter,
                 Phys. Rev. Lett. {\bf 12} (1964) 413.
\bibitem{bahc65} J. N. Bahcall and R. A. Wolf,
                 Phys. Rev. Lett. {\bf 14} (1965)  343.
\bibitem{lati91} J. M. Lattimer, C. J. Pethic, M. Prakash and P. Haensel,
                 Phys. Rev. Lett. {\bf 66} (1991) 2701.
\bibitem{meyer98} E. Chabanat, P. Bonche, P. Haensel, J. Meyer and R.
Schaeffer, Nucl. Phys. A {\bf 635} (1998) 231.
\bibitem{liotta20} N. Sandulescu, Nguyen Van Giai and R. J. Liotta, Phys.
Rev. C {\bf 61} (2000) 061301(R).
\bibitem{estal201} M. Del Estal, M. Centelles, X. Vi\~nas and S. K. Patra,
        Phys. Rev. C {\bf 63} (2001) 044321.
\bibitem{Agrawal05} B. K. Agrawal, S. Shlomo, and V. Kim Au, Phys. Rev. C
  {\bf 72} (2005) 014310.
\bibitem{lala97} G. A. Lalazissis, J. K\"oning and P. Ring, Phys. Rev. C
{\bf 55} (1997) 540.
\bibitem{fsu2} Wei-Chai Chen and J. Piekarewicz, Phys. Rev. C {\bf 90},
(2014) 044305.
\bibitem{chen15} Wei-Chai Chen and J. Piekarewicz, Phys. Lett. B {\bf 748},
(2015) 284.
\bibitem{maza11} X. Roca-Maza, X. Vi\~nas, M. Centelles, P. Ring, and
P. Schuck, Phys. Rev. C {\bf 84} (2011) 054309.
\bibitem{alam15} N. Alam, A. Sulaksono, and B. K. Agrawal,
Phys. Rev. C {\bf 92} (2015) 015804.
\bibitem{biswal15} S. K. Biswal,  S. K. Singh, M. Bhuyan and
        S. K. Patra, Brazilian Journal of Physics, {\bf 45} (2015) 347.
\bibitem{kubis97} S. Kubis and M. Kutschera, Phys. Lett. B {\bf 399} (1997) 191.
\bibitem{nik15} Nik$\check{s}$i\'c, N. Paar, P-G Reinhard and D. Vretnar,
 J. Phys. G {\bf 42} (2015) 034008.
\bibitem{audi12}M. Wang, G. Audi, A. H. Wapstra, F. G. Kondev, M. MacCormick,
        X. Xu and B. Pfeiffer, Chin. Phys. C {\bf 36} (2012) 1603.
\bibitem{angeli13} I. Angeli, K. P. Marinova,
        At. Data and Nucl. Data Tables {\bf 99} (2013) 69.
\bibitem{klup09} P. Kl\'upfel, P. -G. Reinhard, T. J. B\'urvenich, and
J. A. Maruhn, Phys. Rev. C {\bf 79} (2009) 034310.
\bibitem{sharma93} M. M. Sharma, G. A. Lalazissis and P. Ring,  Phys. Lett.
 B {\bf 317} (1993) 9.

\bibitem{thick} A. Trzci\'nska, J. Jastrz\c{e}bski, P. Lubi\'nski, 
F. J. Hartmann, R. Schmidt, T. von Egidy, and B. Klos, Phys. Rev. Lett. {\bf 87} (2001) 082501.

\bibitem{jast04}  J. Jastrz\c{e}bski, A. Trzci\'nska, P. Lubi\'nski, 
B.  Klos,  F. J. Hartmann,  T. von Egidy,  S.  Wycech,  
Int. J. Mod. Phys. E {\bf 13} (2004) 343. 
\bibitem{xavier14} X. Vi\'nas, M. Centelles, X. Roca-Maza and M. Warda,
Eur. Phys. J. A {\bf 50} (2014) 27.
\bibitem{alex10} Alexandros Gezerlis and J. Carlson, Phys. Rev. C {\bf 81}
(2010)  025803.
\bibitem{dutra12} M. Dutra, O. Louren\c{c}o, J. S. S\'a Martins, A. Delfino,
 J. R. Stone and P. D. Stevenson, Phys. Rev. C {\bf 85} (2012) 035201.
\bibitem{daniel02} P. Danielewicz et al., Science {\bf 298} (2002) 1592.
\bibitem{brown00} B. A. Brown, Phy. Rev. Lett. {\bf 85} (2000) 5296.
\bibitem{arumugam04} P. Arumugam, B. K. Sharma, P. K. Sahu,  S. K. Patra,
       Tapas Sil, M. Centelles and X. Vi\~nas, Phys.  Lett. B {\bf 601} (2004)  51.
\bibitem{Haensal17} J. L. Zdunik, M. Fortin and P. Haensel, Astron.
Astrophys. {\bf 599} (2017) 119.
\bibitem{Grill14} Fabrizio Grill, Helena Pais, Constan\c{c}a Provid\^encia, 
 Isaac Vida\~na and Sidney S. Avancini, Phys. Rev. C {\bf 90} (2014) 045803.
\bibitem{Alam17} N. Alam, B. K. Agrawal, M. Fortin, H. Pais, C. Provid\^encia, 
Ad. R. Raduta and A. Sulaksono, Phys. Rev. C {\bf 94} (2016) 052801(R).

\bibitem{ddhd} T. Gaitanos, M. Di Toro, S. Typel, V. Baran, C. Fuchs,
V. Greco and H.H. Wolter, Nucl. Phys. A {\bf 732} (2004) 24.
\bibitem{demo10} P. B. Demorest, T. Pennucci, S. M. Ransom, M. S. E. Roberts,
and J. W. T. Hessels, Nature (London) {\bf 467} (2010) 1081.
\bibitem{antoni13} J. Antoniadis et al., Science {\bf 340} (2013) 6131.

\bibitem{Lattimer13} 
Lattimer and Y. Lim Astrophys. J. {\bf 771} (2013) 51.





\end{thebibliography}
\end{document}